Presented at IDSTA 2026 – 7th International Conference on Intelligent Data Science Technologies and Applications

# Data Quality Assessments: A Theoretically Structured Overview of Approaches and Methods

Ralph Foorthuis
*Digital & Technology, HEINEKEN International*, Amsterdam, The Netherlands
Email: ralph.foorthuis@heineken.com

***Abstract* — The quality of data is crucial for both practice and academia, and this holds for descriptive statistics, AI and advanced analytics alike. This study addresses an apparent gap in the literature and as such presents a full and theoretically grounded overview of data quality assessment approaches and methods, as well as their defining characteristics and interrelationships. For this purpose a broad typology is introduced that employs two theoretical dimensions, namely the *evaluation logic* (*formal* versus *informal*) and the *assessment driver* (*norms* versus *data*). This yields four high-level approaches and 15 methods for assessing data quality. The research results are relevant for academia, as they provide a theory-based overview and definition of the ways that data quality can be evaluated. The study is also relevant for practice because it allows professionals to make informed decisions on using these methods, e.g. as part of an audit or broad data quality assessment strategy.**

## I. Introduction

The quality of data is crucial for both scientific research and industrial practice, and this holds for operational processes, descriptive reporting, advanced analytics and AI alike [1, 2, 3, 4, 5, 6, 7]. Data quality problems have been shown to bring about practical challenges in a variety of domains, including public health [1, 6, 8], the public sector [9], the commercial industry [1, 10] and academia [11]. In *industrial practice* key considerations are typically to manage and standardize the data, and to ensure that the quality is sufficiently high to support and fit fluently with business processes, IT systems and other data [4, 12]. In *academia* the quality of data is crucial to ensure valid research findings and proper management of scientific processes [13, 14]. Low data quality can have a variety of causes, such as data entry mistakes, interpretational differences, suboptimal processes, misalignment and technical errors. In both industry and academia the volumes of data have grown exponentially, in part due to new data technologies, connected devices, and the increasing use and value of artificial intelligence and advanced analytics techniques [5, 15, 22]. These trends also lead to increasing legislation that explicitly demands appropriate data quality [e.g. 16]. Assessing the quality of the used data is therefore a crucial task, not only to gain insight into their current fit for use, but also to identify opportunities for improvement.

However, there are many ways to evaluate the quality of data, each of which has different foci and characteristics. A broad, comprehensive and theoretically grounded overview of such approaches and their defining properties, aimed at a broad group of data management stakeholders, seems to be missing in the literature. This paper therefore offers two contributions. First, a two-dimensional theoretical framework that describes four high-level data quality assessment approaches. Second, an overview of 15 methods for data quality assessment, each of them typified in the framework, described in detail, embedded in the literature, and illustrated with examples from industry and academia. Crucially, the methods are relevant for a broad group of users and practical use cases, defined at a convenient and understandable level of abstraction, and do not only cover technical, automated checks (see sections II and IV for more on this).

This paper proceeds as follows. Section II discusses the literature, the key concepts and the two-dimensional theoretical framework that describes four data quality assessment approaches. Section III describes these general assessment approaches in more detail, as well as the more specific methods that are positioned within the framework. Sections IV and V are for discussions and conclusions respectively.

## II. Literature and Theory

### A. Related Work

A substantial number of publications provide overviews of crucial data quality (DQ) concepts. These include overviews of DQ dimensions [1, 3, 18, 19, 20], DQ problems [6, 17, 24], data anomalies [21] and prevention and improvement techniques [3, 17, 24]. However, none of these studies provides the broad overview of high-level DQ assessment approaches or concrete methods as defined in this study.

Studies [22, 24] do provide overviews of detailed techniques (“algorithms” or “low-level functionalities”) that represent DQ assessment checks, such as “string values are of right length”, “row satisfies SQL expression” and “primary/ foreign key analysis”. However, compared to the methods described in this study these reside at a significantly more detailed level of abstraction and cover a limited scope (only automated checks, but not e.g. interviews). The same holds for the detailed statistical checks discussed in [23] and the metrics in [2, 3]. Conversely, the main overview presented in [3] sits at the very abstract end of the spectrum, covering not concrete techniques but entire methodologies such as “Total Data Quality Management” and “Canadian Institute for Health Information methodology”. DAMA [20], despite also targeting a broad group of data management stakeholders, does not provide a comprehensive overview of methods either. To conclude, as far as I am aware the theoretically structured overview presented here is not published yet. This study offers a wide overview of methods, and in such a way that these are relevant for all data (quality) stakeholders (e.g. not detailed techniques that are only relevant for technical specialists). See section IV for more on this topic.

### B. Definitions and Concepts

*Data quality* is defined in this study as the extent to which data are fit for use by consumers, both for specific, known use cases and for a wide variety of as-of-yet unknown future use cases [cf. 1, 4, 18]. This emphasizes that data quality is not only associated with *specific* requirements, which only hold for certain use cases and contexts, but also with *generic* expectations, which imply norms that in principle always hold and are independent of the use context, and thus allow repurposing of datasets [cf. 1, 19]. An example of a specific requirement is the timeliness of the data, which strongly depends on the given business process (e.g. a certain BI report may well be delivered monthly, while a security event should typically be processed as soon as possible to limit the potential damage). An example of a generic requirement is the format conformance of data values to their predefined data types (e.g. having valid timestamps in a date-time attribute and quantitative values in a numerical attribute).

There is a large body of research on *data quality dimensions*, which represent the different aspects on which the quality is evaluated. The literature shows overlapping – but not identical – views on the dimensions that need to be taken into account [c.f. 1, 3, 14, 19, 20]. The current study uses the "common dimensions" of DQ identified in DAMA [20]: completeness, consistency, accuracy, format conformance (including validity and referential integrity), reasonability, uniqueness and timeliness.

A *data quality assessment* can be defined as an evaluation of data quality on these dimensions [15]. The methods that prescribe how to do this are the focus of this study.

### C. Theoretical Framework

The data quality approaches are distinguished and described by a framework using two dimensions: evaluation logic and assessment driver.

- ***Evaluation Logic***: Whether the logic used to evaluate the data and arrive at conclusions regarding their quality is formal or informal [25, 26, 27].

  *Formal logic* employs an exact premise-conclusion form, i.e. strict pre-established rules to draw conclusions via deductive inference. This is the logic of mathematics and programming languages that is used to specify algorithms that do not allow different interpretations. It is well-suited for automation and large-scale repetition with deterministic and consistent outcomes. However, it may sometimes be difficult to employ this logic meaningfully in a practical setting due to its rigidness and the fact that the rich real-life context is often ignored.

  *Informal logic* employs critical thinking, natural language, and the interpretation and evaluation of arguments and norms. This is a more flexible logic and well-suited for a concrete, meaningful and practical context, but it is not easy to automate or consistently repeat.

  Note that this dimension pertains to the *logic* used, not to the role of the assessment in the overall process. Therefore, an audit using only informal logic can still draw 'formal' (in the sense of 'official') conclusions for the organization.

- ***Assessment Driver***: The main driver and orientation that determine the design and execution of the data quality assessment. An assessment can be norm-driven or data-driven [21, 28, 29].

  *Norm-driven* means that the design and execution of the assessment are determined by *predefined rules, principles, specifications or other types of norms*. This requires sufficient preparation before executing the evaluation.

| | | Assessment Driver | |
|---|---|---|---|
| | | **Norms** | **Data** |
| **Evaluation Logic** | **Formal** | Specification-Based Validation<br>Strict, deterministic, executable domain-based validation rules and structures are used to detect erroneous or suspicious data.<br>Number of methods: 4 | Data-Based Comparison<br>Strict, deterministic comparisons of available data are used to detect unwarranted inconsistencies or duplications across situations and datasets.<br>Number of methods: 3 |
| | **Informal** | Perception-Based Validation<br>Non-strict, non-deterministic methods are used – often in a non-automated way – to evaluate the data on predefined norms.<br>Number of methods: 4 | Exploratory Data Quality Analysis<br>Data-driven and exploratory analysis methods – without strict predefined quality expectations – are used to detect unlikely and suspicious data.<br>Number of methods: 4 |

Fig. 1. Typology of data quality assessment approaches

  *Data-driven* entails that the *data themselves*, rather than pre-defined norms, function as the main determinants of the design and execution of the DQ assessment. This allows making opportunistic use of data availability, a fast start of the assessment, and an exploratory process in which the data speak more for themselves and new DQ questions can be iteratively added to the analysis. (See section IV for more on norm- vs. data-driven.)

This yields a typology that distinguishes between four fundamentally different data quality assessment approaches (quadrants), as shown in Fig. 1. The next section discusses the approaches and their methods in detail. See Fig. 2 for the full typology with approaches as well as methods (and see Appendix A for the large reader-friendly version).

## III. Data Quality Assessment Approaches and Methods

This section typifies and describes the four approaches and 15 methods for data quality evaluation.

### A. Specification-Based Validation

In this assessment approach the domain's requirements form the basis for formal and executable validation rules and structures that are used to detect erroneous or suspicious data. These testable specifications may first manifest themselves in the form of *data models* and implemented *schemas*, for example as the attributes (i.e. the fields, variables or columns) and their data types, such as integer, real, string and categorical enumeration, as well as whether attribute values are optional or mandatory [30, 31]. The data model also specifies which entities are related to each other, and what the cardinality of each relationship is (i.e. defining referential integrity). This declares, for example, that a natural person can have zero, one or multiple children, or that a sales order can have one or multiple line items. Beyond this structural metadata, *domain-based rules* can be used to specify restrictions on the values within and between attributes [20, 22, 24]. For example, a categorical attribute typically has a known list of allowed class values (enumeration or lookup table). If this is taken further, then a dictionary can be used to check if the words used in unstructured text are valid [24]. Similarly, a numerical attribute may come with a specification of its value domain, e.g. that it needs to be an integer between 5 and 25. More complex rules can also specify allowed value combinations between multiple attributes or records. An example of such a rule is that a person under the age of 18 is not allowed to have a driving license.

A crucial benefit is that these data models and domain rules are defined in a formal, deterministic logic and can therefore be translated into machine executable specifications that reliably and consistently verify the quality of large volumes of data. Database management systems and integration platforms (middleware) typically come with automated parsing and validation functionality to evaluate whether the actual data comply with the structural aspects of the predefined data model or schema. This allows detecting invalid data types (e.g. a textual character in a numerical attribute) and relationships (e.g. repeated values while the cardinality is defined as one). Allowed values in a single attribute as well as allowed value combinations of multiple attributes are often verified using predefined lists and tables that hold the valid values and their combinations. Other domain-based rules manifest themselves as IF THEN ELSE statements that specify restrictions within or between attributes, such as for the previously mentioned example: **IF Person.Age < 18 THEN Person.Driving_License = FALSE**

Predefined expectations can also pertain to the entire dataset. A variety of *distributional tests* allow for validating the dispersion of the collected data. If the population distribution is known upfront it can be statistically tested whether the data at hand are consistent with the expectation [22, 23, 32]. Various Goodness-of-Fit tests allow for verifying whether the data come from a given distribution. The predefined distribution can be theoretical (e.g. a Gaussian, Binomial or Log-Normal distribution) or may be evaluated by means of subtotals based on previously collected data (e.g. using a Chi-square test). Especially for statistical and AI use cases it is crucial to evaluate the representativeness, class (im)balance, noise, drift and other forms of bias [23, 52].

Finally, *supervised anomaly detection* and *related advanced analytics* are also specification-based methods. These allow finding deviations by using known and predefined patterns, but will be discussed in section III.C.

Specification-based assessments are regularly performed in academic research to check the quality of the collected data. For example, [11] uses various rules to verify the validity of participants in an empirical study, such as having more than 4 children of the same age. Several rule-based checks are used in [12] to assess the quality of survey responses. One of these checks assessed whether there was a basic consistency between the responses of questions that were content-wise related. In addition, rule-based matching (record linkage) was used to prevent duplicate submissions. Commercial and governmental organizations also heavily employ specification-based data quality validations [20, 31].

In addition to their suitability for automation and scaling, a crucial benefit of data quality specifications is that they can often be executed at the time of data entry, thereby not only verifying but also directly enforcing the quality while recording the data. The specification-based quality approach can thus also function as a preventative measure, unlike the other three approaches, which only offer an after-the-fact data quality evaluation.

A wide variety of data quality dimensions are relevant for this assessment approach. In particular *format conformance* (i.e. the data need to be compliant with the predefined data structures, relations and value domains), *consistency* (rules can involve restrictions on the combination of attribute values) as well as *completeness* and *uniqueness* (specifications can be used to identify missing or duplicate data). Finally, distributional tests can evaluate the *reasonability* (e.g. representativeness, class balance and drift).

### *B. Data-Based Comparison*

This approach starts from the available data and verifies their uniqueness and consistency if they are expected to be unique, identical or at least not conflicting across datasets or situations. This needs to be evaluated because unwarranted duplicates can exist within or across datasets [24]. Conversely, unwanted differences may occur in intentionally duplicated datasets [1, 3, 15]. External reference data, such as ISO codes, national street registers and other governmental data, are regularly present inside multiple systems as lookup tables [3, 19]. The organization's internally produced data are often also present in more than one system. Data are ideally not duplicated because of the costs of replication and the risks of inconsistency. However, in practice systems often demand that the data are stored inside their own databases for convenient processing, or as a result of high stability and performance requirements. Moreover, crucial data may be maintained in a master data management system before they are shared with multiple business applications, which in their turn offload data to a separate data and analytics platform. Copying data is not always as self-evident as it may seem at first glance. Data replication can come with highly complex update logic and may involve manual steps. Replication thus comes with the realistic risk that two independently replicated datasets are not identical and differ in unexpected ways from the original source.

Comparing data values across systems and datasets by using clear logic, possibly in batch mode, is therefore an important real-life data quality use case. A data-based comparison verifies whether data that should be identical or consistent are indeed so. Relatedly, it verifies whether data are not accidentally duplicated where they are expected to be unique. Such an assessment is typically realized by means of a straightforward *comparison of attribute values*, but could also require some translation logic if the data representation is different across the datasets (note that matching logic belongs to a specification-based norm in III.A). Moreover, the focus may be on *individual instances*, but may also involve comparing the *total number of instances* or *sub-group totals*.

Alternatively, a summarizing *digest*, such as a cryptographic checksum, can be generated of each dataset, or parts thereof, and compared to verify whether the data or files are identical [33, 34]. The slightest difference will result in different checksum values. Simpler variants are also possible, such as comparing file sizes. A digest verification is particularly relevant for guaranteeing the integrity of data transfers and storage. Crucially, this can be done without the need for transporting the data and without understanding the schemas of the datasets. However, the downside is that such digests essentially treat the data as an unstructured black box (e.g. do not take into account the data model) and thus do not provide insights into internal data quality aspects.

In addition to comparing *existing (registered) data*, the internally available data can be compared to (newly captured data from) *the real world*. For example, in the context of a DQ audit a sample of customers may be (re)observed in the real world, registered according to the protocols and specifications, and subsequently compared with the existing data [18]. Similarly, a service desk agent may ask certain data-validating questions during interactions with the customer. The check can also be simulated in order to verify whether the recording process correctly captures the data from the real world into the system. This is done, for example, as part of data quality audits in the health sector [35].

Similar to the *specification-based assessment* discussed in the previous section, an important benefit of the data-based comparison is that it is formal in nature and therefore automatable and scalable. The comparisons can be specified in a programming language for large-scale automated evaluations with deterministic and repeatable outcomes.

An example of a research paper using comparisons of data from different sources as part of a quality audit is [36], in which a sample of cancer patient data is compared across two medical registries to verify their completeness and accuracy. The identified discrepancies could not be directly remedied because neither of the two systems could act as a golden source, but were corrected during the next patient visit.

Another example can be found in [37], in which the frequency of received questionnaires per economic sector is compared with similar data from previous research [38] in order to verify the representativeness of the sample. This same check is performed later again in [39] with a third sample. Note that at some moment the results of such an analysis may yield concrete expectations that can be formally specified without the need for data comparisons. This results in a specification-based approach as described in section III.A, namely a *distributional test.*

This approach touches various DQ dimensions. The data-based approach contributes mostly to checking whether the data are *consistent* or *unique* across different systems, copies or variations. The *completeness* dimension is also relevant because data may have been incompletely replicated or collected, which can be discovered during the comparison. Finally, if one of the compared datasets (e.g. an internal master data management system or an official external data register) or the newly gathered data can be regarded as representing the truth, then *accuracy* can also be verified.

### *C. Exploratory Data Quality Analysis*

This approach aims to identify suspicious and unlikely data in an exploratory, data-driven fashion without using predefined quality specifications or strict expectations. Although this approach may employ formal algorithms to detect deviant cases, the final decision on the quality needs to be taken using informal logic because unusual cases do not necessarily reflect bad data quality – the world may simply allow anomalous but correct instances [21]. An exploratory analysis is valuable because it yields insights into new datasets and enriches the knowledge about existing sets, including possible errors [28]. It facilitates discovering hitherto unknown data quality issues and other insights because the narrow-mindedness that comes with predefined specifications is avoided.

One major exploratory method is *unsupervised anomaly detection*, which covers a broad range of data analysis algorithms and techniques that identify deviations from the common patterns that may indicate suspicious or interesting instances. Other names for (variations of) such an analysis are outlier analysis, deviation detection, novelty detection, and out-of-distribution detection. The aim is to find data instances that are both *rare* and *different* and therefore *do not fit the general patterns* exhibited by the majority of the data [21, 40]. In order to detect such anomalies it is not required to have predefined specifications or definitions of these deviations because an algorithm can determine what the normal patterns are and subsequently which cases are abnormal. Domain knowledge is also not needed because (non-parametric) algorithms will discover these normal patterns without many assumptions regarding the distribution of the data. Finally, anomaly detection is not only applicable to structured data, as many algorithms for the analysis of texts, images, video and audio are also available.

To facilitate the discovery process and interpret the results it is worthwhile to use an anomaly framework that provides an overview of known anomaly types and explains why these are rare and different in terms of fundamental data dimensions. A literature study of 250 years of anomaly detection research yielded a comprehensive overview of the concrete (sub)types of deviations that one can encounter in data [21]. The framework indicates which kinds of anomalies may be present in datasets, depending on various fundamental dimensions of data, including the *data structure*, the *data types* of the attributes, and the *data distribution*. It also indicates other defining properties of the anomaly, such as whether it is *atomic* (i.e. a single data point) or *aggregate* (i.e. a group of data points), and whether it is *univariate* or dependent on *multivariate* associations between attributes. The framework’s tangibly defined anomaly subtypes guide the analyst in what deviations to search for when conducting an exploratory analysis of a certain kind of data collection. See Appendix B for more on this framework.

An exploratory anomaly analysis may yield new insights into data issues, and subsequently drive the development of detection rules and model specifications as described in the approach of section III.A. It can also facilitate solid testing of IT systems and processes by exposing them to low quality data that contains various types of deviations.

An example of employing unsupervised anomaly detection for data quality purposes can be found in [9]. In this study the SECODA algorithm was used to identify unusual data points in a governmental database with both numerical and categorical attributes. The analysis detected various types of anomalies, and some of the cases proved to be indicative of previously unknown quality problems resulting from the data export software. This exploratory analysis led to the software being updated with a more sophisticated export function.

The above refers to *unsupervised* anomaly detection, which does not require knowledge of the anomalies beforehand. Anomalies are detected by first discovering the normal patterns in the data and subsequently identifying the deviations – predefined specifications are thus not required. However, anomaly detection can also be *supervised*, which uses domain knowledge, rules, known distributional assumptions and/or prior labelling and training of a statistical or AI model. Because this detection process essentially requires upfront definitions of (ab)normality this sort of anomaly detection should be regarded as a form of specification-based data quality assessment, as discussed in section III.A.

Apart from unsupervised anomaly detection, other forms of exploratory data quality analyses also exist. A manual *exploratory data analysis* (EDA) aims at describing data as well as identifying patterns and deviations [20, 28, 41].

Rather than depending heavily on algorithms (as with unsupervised anomaly detection described above), the emphasis here is on manual exploration. One technique is to *visualize* the data. A simple scatter plot may readily point to extreme tail values, peripheral points, and unusual classes [21, 28]. Visualizations such as box plots and histograms also help to identify anomalies [28, 41, 42]. Another technique is *manually querying* the data. For example, a wide variety of SQL statements can be used in an exploratory analysis to discover extreme cases, unusual classes, orphaned records, null values, duplicate instances and many other suspects [30].

One can also employ *automatic profiling* of a dataset to determine its metadata and discover properties of the data therein [10, 14, 43, 51]. Profilers are typically able to identify attributes and their technical data types, often-used functional data types (e.g. dates, postal codes, credit-card information, latitude-longitude coordinates), relationships, inconsistent formatting and missing values [10, 19, 20, 42, 43]. Several profilers also come with out-of-the-box visualization, duplication analysis, and anomaly detection capabilities, thereby automating some of the tasks described above [14, 19, 42]. Automatic data profiling is therefore a method that is accessible to a broader range of users than manual exploratory data analysis and unsupervised anomaly detection, both of which require more advanced skills.

| | | **Assessment Driver** | |
|---|---|---|---|
| | | **Norms** | **Data** |
| **Evaluation Logic** | **Formal** | Specification-Based Validation<br>• Data model and schema validations<br>• Executable domain-based rules<br>• Distributional tests<br>• Supervised anomaly detection and related advanced analytics | Data-Based Comparison<br>• Comparisons of data values or (sub)totals between datasets<br>• Comparisons of digests (e.g. checksums) of different datasets or selections<br>• Comparisons of data values or (sub)totals with reality or simulations |
| | **Informal** | Perception-Based Validation<br>• Interviews, focus group discussions and related methods<br>• Desk reviews of documents, (un)-structured data and other artifacts<br>• Field observations<br>• Questionnaires | Exploratory Data Quality Analysis<br>• Unsupervised anomaly detection<br>• Manual exploratory data analysis<br>• Automatic data profiling<br>• Large language models (LLMs) / generative AI (GenAI) |

Fig. 2. Approaches and methods (see Appendix A for the large version)

Finally, *large language models* (LLMs) and *generative AI* (GenAI) also provide functionality to verify data quality. In particular, in case of missing or unusual text values and their formatting, LLMs can flag the respective records or sections and even impute the values that are expected to be correct [44]. Also, due to their reasoning capabilities LLMs may be applicable in situations that involve background data, large unstructured data sections, and subjective decisions [45]. Note that, although the underlying algorithms are formal in nature, the output of the trained statistical language models – with their settings (e.g. temperature) and billions of parameters – cannot be regarded as being fully predictable and based on deterministic rules [46, 47]. In addition, the results prove to be highly sensitive to the prompts provided [45]. Some form of human oversight is therefore still advised [ibid]. Consequently, this type of analysis is positioned in the exploratory data quality analysis quadrant. However, further developments in this area may lead to techniques that are capable enough to be included in another quadrant. Also, in a more complex agentic setup an LLM-based agent may call a deterministic tool that uses formal logic to execute strict data quality validations, in which case this part of the validation should be regarded as a specification-based validation or data-based comparison.

Various data quality dimensions are relevant for this section's assessment approach. An exploratory analysis directly contributes to the *reasonability* dimension, with the identified deviations questioning the quality without being able to immediately draw definitive conclusions. The *consistency* dimension is also relevant here, as anomalies are by definition inconsistent, i.e. not conforming to the overall patterns in the dataset.

### D. Perception-Based Validation

The assessment methods in this quadrant use informal logic and therefore do not employ strict, deterministic rules that can be automated in a straightforward and easily repeatable fashion. However, data quality is evaluated using norms that are established before the assessment is conducted. These may be detailed and specific rules or more generic and abstract principles, depending on the situation.

Unlike those in the other three approaches, the methods in this quadrant often cannot analyze large volumes of granular data in sets or systems, but rather aim directly at the data quality at a higher level of abstraction, as experienced by data stakeholders. This typically involves qualitative assessment methods such as *interviews* and *focus group discussions* with knowledgeable informants, *desk reviews* of relevant documentation and other artifacts, as well as *field observations* of data and processes in action [3, 18, 20]. Desk reviews allow for a wide variety of verifications, such as manuals, designs, standards, logs, previous evaluation reports, but also data (e.g. on paper). Moreover, although some aspects may be automatically verified by means of a specification-based method, a desk review is the typical method to evaluate whether the schemas of implemented datasets and message structures comply with the organization-wide canonical data models and standards. Quantitative assessment methods can also be employed in this approach, in particular *questionnaires* that allow for creating statistical overviews of stakeholder perceptions [3, 18].

These methods can be used for various purposes. First, they help to identify shortcomings in the data as perceived by users. A benefit of subjective assessment methods is that they discover *relevant* data quality problems, since they are informed by the stakeholders that actually use the data and are able to comment on the real and practical problems they face in their daily work activities. Second, they can be used to determine whether alleged errors – discovered via e.g. unsupervised anomaly detection – are *true errors or false positives*. Drawing upon their domain knowledge the interviewed stakeholders can make appropriate judgments on the data quality of suspicious cases.

Finally, these methods are able to explain *why* the confirmed true errors occur, i.e. how and under what circumstances they are introduced into the data. Especially qualitative methods – such as interviews, focus groups and desk reviews – are able to take into account the complex context and identify the root causes of problems, and yield insights beyond the mere level of data quality. They can explain and enrich the results of the assessment methods in the other quadrants, which target only the data quality and not the underlying causes and context. For example, a specification-based assessment method can provide exact percentages of certain errors, but interviews with relevant stakeholders are often the most effective way to understand why these errors are present and how they come about in the dataset. This may also give concrete pointers to prevent low data quality in the future, for example by improving the processes and IT systems that capture, create and use the data.

Examples of the use of perception-based methods can be found in [48], which assessed the data quality of three hospitals in the Nairobi metropolitan area of Kenya. The researchers used workplace walkthroughs (field observations) to observe and learn the daily practices of the nurses, doctors, laboratory technologists, pharmacists, cashiers, medical record officers and other roles, with a particular focus on their data-related activities regarding the creation, transmission and usage of health information. Structured interviews with the above-mentioned stakeholders were subsequently held to collect inside views on the data quality experienced in several processes. Finally, actual data records (both analogue and digital) were analyzed w.r.t. completeness and accuracy. The combination of these methods resulted in a broader view on the data quality and yielded suggestions for improvement.

In principle all data quality dimensions can be relevant for a perception-based assessment. In fact, before the interviews, document analyses or field observations take place, the list of dimensions is typically used to scope and focus the questions that need answering. In the context of a data quality audit they may shape the audit norms. Perception-based methods can also investigate dimensions that are difficult to assess otherwise, such as the *timeliness* of the data for real-life tasks.

## IV. Discussion

This section discusses several less obvious differences and relations between the approaches. In particular, it is worthwhile to note that the *specification-based validation* and the *exploratory data quality analysis* are *fundamentally different* regarding their design, execution and value. The former is driven by internal standards, the employed data formats, and external laws and regulations. These are captured in formal specifications, i.e. schemas and rules, that facilitate strict compliance checks that manifest themselves as validations on entities, attributes, and complex relationships between multiple attributes. Because they encode formal norms on how the data should be, they yield conclusions on data quality that are (more or less) 100% true. Not only the checks themselves are therefore well-suited for automation, but the process of drawing conclusions and possibly of corrections and other follow-up actions as well. This can obviously be very valuable, especially for compliance purposes. Nonetheless, it is good to note that this approach is also somewhat tautological in nature. We first specify that an attribute can only contain the values “A” and “B”, and if we subsequently find a “C” we conclude that it is wrong. Not a very surprising outcome! This approach will therefore not yield new ideas or deeper insights about the data.

This is very different with an *exploratory data quality analysis*, which is based on algorithmic or manual exploration without (the need for) predefined norms for the data under scrutiny. Similar to the *specification-based validation* the findings can pertain to entities, single attributes, multivariate inconsistencies or the dataset as a whole. The key point is that no a priori norms or knowledge of the domain or dataset are required beforehand, as the aim is to let the data speak for themselves. As opposed to the *specification-based* approach this analysis can start immediately after obtaining the data. However, the results are also more uncertain, because the identified anomalies are not guaranteed to represent erroneous or interesting cases. They might simply be statistical flukes or correct measurements that reflect a real world that is more complex and diverse than assumed. The decision-making process following the analysis therefore cannot be simply automated because this requires domain knowledge and a conscious follow-up analysis to draw definitive conclusions about the true meaning of the deviations and whether further actions are required. Moreover, the results may depend on the idiosyncrasies of the specific analysis techniques and algorithms, as these will all yield a different set of identified deviations [49, 50]. In spite of that, another valuable aspect of this approach is that it helps discover unexpected and surprising phenomena, and facilitates creating and enriching the domain knowledge.

In short, predefined rules, schemas and other specifications work well for compliance purposes and the automation of large parts of the data quality process, while unsupervised anomaly detection and related exploratory techniques work well for detecting implausible cases, enriching knowledge, and possibly developing new rules and specifications.

Another point worth discussing is that the approaches and methods discussed in this paper can be *combined* for a richer data quality assessment. Each has its own merits and focuses on different aspects, and depending on the situation it may be beneficial or even necessary to combine several approaches and methods. For example, an *exploratory quality analysis* may bring anomalous and suspicious data to light. However, that analysis is performed by data scientists who excel in querying datasets and employing algorithms to detect deviations from the overall data patterns, but who may have no deep domain knowledge of the data or the processes that create and use them. If that knowledge is not readily available, for example because it is spread across the organization, then a *perception-based approach* can subsequently be employed to understand whether the suspicious cases are either faulty instances or correct data that are manifestations of a noisy and complex world. This may involve interviewing or observing professionals who have inside knowledge of the relevant processes, systems and practical situations. In addition, documents and other artifacts can be studied in detail to deepen the insights. Moreover, the analysis may conclude that certain anomalies indeed point to true quality issues in the dataset. These insights can then lead to formulating formal rules and schemas to validate new data by means of a *specification-based validation* in the regular processes that create and collect the data.

Study [48] indeed found that combining different methods in their hospital study allowed for a broader and richer assessment of data quality, and that this yielded suggestions for improving the processes and systems. Moreover, they concluded that the ‘meta quality’ of the assessment increased due to the fact that a mixed-methods approach made it possible to corroborate the outcomes. In other words, the assessment results were more robust because one method’s findings could be used to verify and correct the findings of another method.

Table presents 1 an overview of characteristics that are unrelated to the four approaches. It shows that many methods can deal with both structured and unstructured data. Only checksums and related digests deal purely with unstructured data (by ignoring the data model). The table also indicates which methods allow an exploratory view on the data. Note in this regard that some norm-based methods also offer room for exploration if the opportunity arises, e.g. while conducting an interview. Finally, the table shows which methods allow taking input not only from the available datasets but also from the external, real world.

| Data quality assessment method | Data structures | Allows exploratory investigation | Uses input from ext. reality |
|---|---|---|---|
| Data model & schema validation | Structured | No | No |
| Executable domain-based rules | Both | No | No |
| Distributional tests | Both | No | No |
| Supervised anomaly detection & related analytics | Both | No | No |
| Comparison of data values and/or subtotals | Structured | No | No |
| Comparison of digests (e.g. checksums, file size) | Unstructured | No | No |
| Comparison of data with reality | Both | No | Yes |
| Unsupervised anomaly detection | Both | Yes | No |
| Manual exploratory data analysis | Both | Yes | No |
| Automatic data profiling | Both | Yes | No |
| Large Language Models / Generative AI | Both | Yes | No |
| Interview, focus groups & related methods | Both | Yes | Yes |
| Desk reviews of (un)structured data & other artifacts | Both | Yes | Yes |
| Field observations | Both | Yes | Yes |
| Questionnaires | Both | No | Yes |

Table 1. DQ methods and additional characteristics

Despite the fact that the *data-based comparison* and *exploratory DQ analysis* are data-driven approaches, norms or metrics can still play a role. After all, an assessment implies evaluation on certain dimensions. However, in the exploratory analysis of section III.C strict norms will not be present, especially not upfront. Anomaly detection often results in cases getting assigned an anomaly score, and depending on the available time and resources analysts will scrutinize the most extreme deviations, independent of a pre-defined threshold for distinguishing 'normal' from 'anomalous'. Moreover, even extreme anomalies may not be true data errors because reality allows strange but correct cases. For these reasons an exploratory analysis may iteratively result in – not start with – clearly defined norms for (in)correct data. In the data-based comparison of section III.B strict assessment logic does get used. However, this is often straightforward and limited, because the main verification is simply testing whether the data are identical or consistent. Also, the setup of the assessment may be more influenced by the crucial datasets that need to be evaluated than by whether clear predefined norms exist.

It is valuable to return to the topic of how DQ methods are defined here. They are scoped in such a way that they are understandable and usable for a broad group of data quality stakeholders in a real-life (academic or industrial) setting. This has several concrete implications. First, methods are defined at a level of abstraction that yields a practical 'toolbox' (Fig. 2) that features a comprehensive and comprehensible set of methods. These can be picked for an audit or other short-term DQ analysis, or may be included in a long-term DQ strategy. Second, the methods are not primarily defined in terms of functional results ('what'), but in terms of the work that needs to be done ('how'). For example, one of the methods in this study's typology is *comparison of data values*, which focuses on the work or task (instead of focusing on the functional results, which in this case are the identification of *inconsistent data* and of *duplicate data*). This type of method definition makes the typology practically usable, since an individual method is not a complex combination of many (possibly very different) *skills, knowledge and resources*. In the exploratory data quality analysis quadrant, therefore, *automatic data profiling* and *manual exploratory data analysis* are distinct methods, despite the fact that these can come up with similar findings. However, they require different skills and knowledge, with the latter demanding more analysis and engineering expertise.

The results of this research study are relevant for both academia and industry. The paper presents a structured typology for data quality approaches and methods, thereby contributing to academic theory on data quality and analysis. Moreover, scientists can use the typology and the overview of approaches and methods therein to decide on the most appropriate methods for verifying the empirical data they have collected in their research projects. This study's results are also highly relevant for industrial practice as organizations can use the typology to pick and choose data quality methods for specific use cases or to develop a broader data quality strategy or audit.

## V. Conclusion

This study has presented an overview of 15 data quality assessment methods. By means of a structured typology that employs two dimensions these methods are typified by four fundamental quality assessment approaches (see Fig. 2). In addition, the approaches are illustrated by examples from both academia and industrial practice.

This broad overview of methods has, as far as I am aware, not been published before. In addition, this study introduces a theoretical framework to typify and discuss them. The results are relevant for both practice and academia.

Future research can evaluate the typology in an empirical setting. As part of this it can be investigated how the different methods can be combined to arrive at a consistent and effective overall data quality strategy or audit. Finally, future studies can focus on identifying or designing additional assessment methods and positioning them in the framework.

## Notes

(Gen)AI functionality has been used for simple DQ tests with the LLM method, not to generate texts, visuals or ideas.

### Appendix A: Overview of Data Quality Approaches and Methods

The figure below is an enlarged and enriched version of Fig. 2, i.e. the typology with DQ approaches and methods.

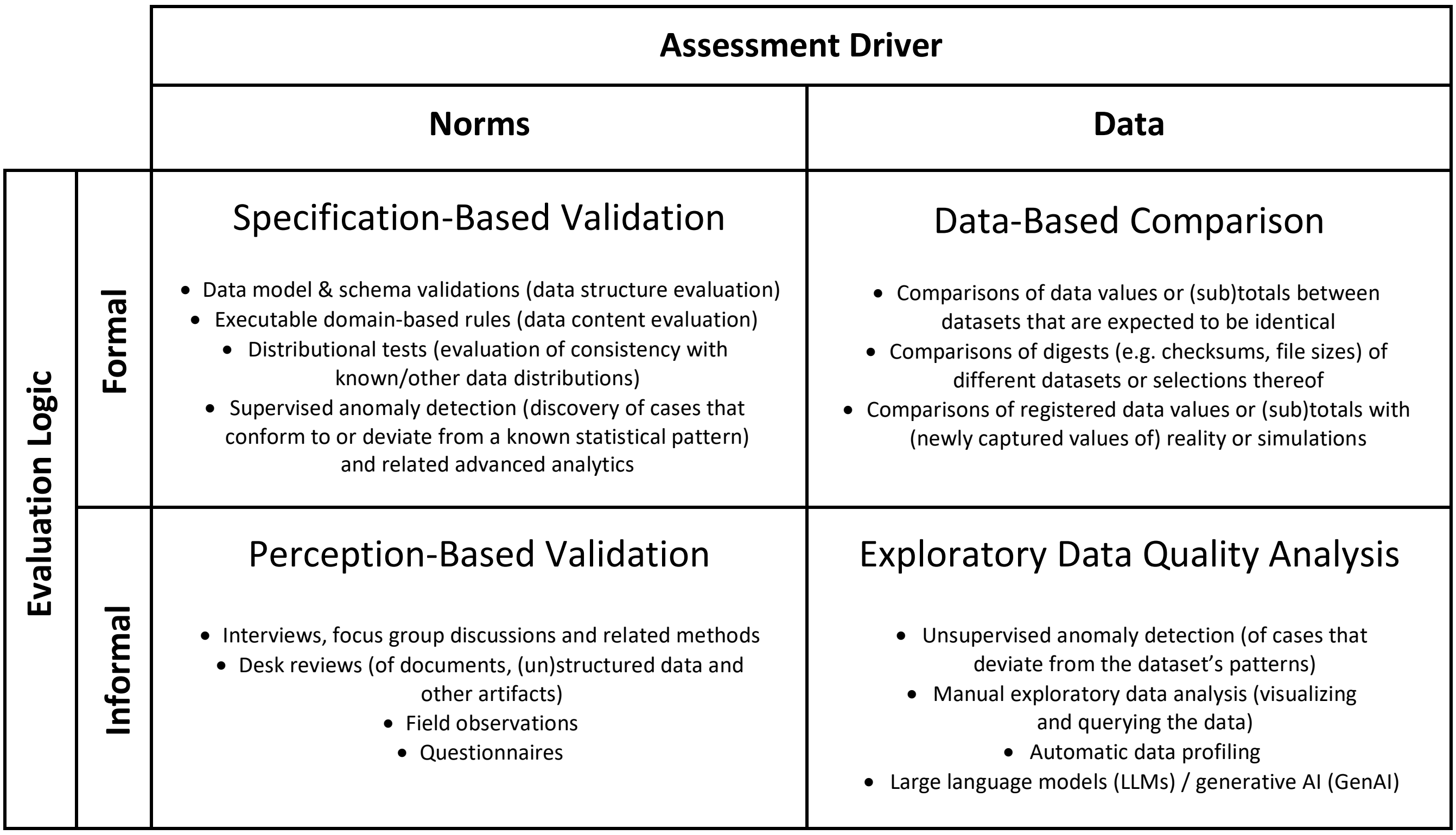

### Appendix B: Overview of Deviations in Data

The diagram below presents an overview of anomalies that can be detected via an Exploratory Data Quality Analysis approach (see section III.C). Anomalies are deviations that may be encountered in datasets and that may point to data quality issues or to interesting occurrences. The typology from [21] uses five fundamental data dimensions, resulting in the definition of 3 anomaly super types, 9 basic types, and over 60 concretely defined subtypes.

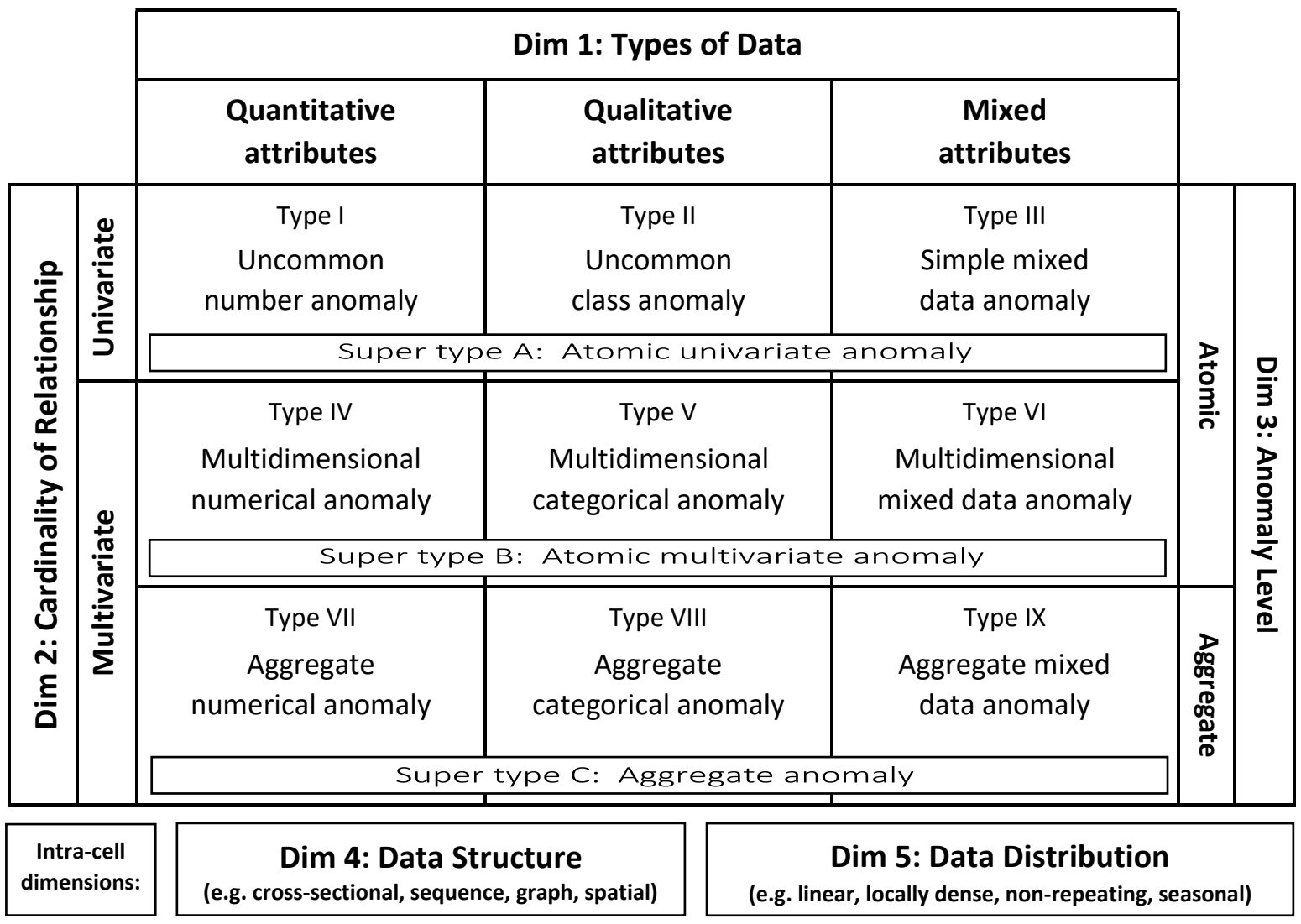

The concrete subtypes are not shown here, but some examples are discussed below. One Type I anomaly is the *extreme tail value*, which is the classical outlier, a suspiciously high or low numerical value. A Type II example is the *deviant repeater*, which is a categorical value that occurs relatively frequently while the normal values do not repeat themselves. This anomaly can point to problems with identifiers or other fields that should be unique. The *peripheral point* is one of the Type IV anomalies and takes the form of an isolated instance that lies outside the relatively dense multivariate clusters without exhibiting extreme values, and may indicate faulty data entries. Finally, one of the Type VI subtypes is the *incongruous common class*, which has a categorical value that is normal, though not in that area of the numerical data space. This may be caused by either a wrong class label or erroneous numerical values. The full typology describes 63 of such concrete anomaly subtypes that can be searched for in an exploratory analysis.

See the video *Taming the anomaly: An overview of outliers and other deviations in data* for concrete and animated illustrations of a large number of anomalies (https://www.youtube.com/watch?v=XOBR2sc00y4). Example datasets and R code with algorithms and anomaly detection examples can be downloaded from Github (https://github.com/ralfoan).